%% file: lsst.tex
\documentclass[11pt]{article}
\usepackage{amsmath,amsfonts,amsthm,epsfig}
\usepackage{fancybox}

\input{prestuff}

\def\Reals#1{\mathbb{R}^{#1}}

\def\vs#1#2#3{#1_{#2},\ldots , #1_{#3}}

\def\union{\cup}

\def\Union{\bigcup}

\def\bdry#1{\partial \left(#1 \right)}

\def\edgbet#1{E \left(#1 \right)}

\def\vol#1{\mathrm{{vol}}\left(#1  \right)}

\def\Gradius#1#2{\mathrm{rad}_{#1}\left(#2  \right)}
\def\cost#1{\mbox{cost}\left(#1  \right)}

\def\expec#1#2{\mbox{\bf E}_{#1}\left[ #2 \right]}

\def\norm#1{\left\| #1 \right\|}

\def\gennorm#1#2{\left\| #2 \right\|_{#1}}

\def\setof#1{\left\{#1  \right\}}
\def\sizeof#1{\left|#1  \right|}
\def\abs#1{\left|#1  \right|}

\def\calL{\mathcal{L}}
\def\Gdist#1#2#3{\textrm{dist}_{#1} (#2, #3)}
\def\dist#1#2{\textrm{dist} (#1, #2)}
\def\tdist#1#2{\textrm{dist}_{#1} (#2)}
\def\stretch#1#2{\textrm{stretch}_{#1} (#2)}
\def\avestretch#1#2{\textrm{ave-stretch}_{#1} (#2)}

\newcommand{\BallShell}[2]{\ensuremath{{\mathop{\mathrm{BS}}\nolimits}({#1},
{#2})}}
\newcommand{\AS}[1]{\ensuremath{{\mathop{\mathrm{AS}}\nolimits}
({#1})}}
\newcommand{\TS}[1]{\ensuremath{{\mathop{\mathrm{TS}}\nolimits}
({#1})}}
\newcommand{\GraphSequence}[1]{\ensuremath{\sigma({#1})}}
\newcommand{\Contracted}[1]{\ensuremath{\widetilde{#1}}}
\newcommand{\OriginalVertices}[1]{\ensuremath{\chi ({#1})}}
\newcommand{\AvoidOverfull}[0]{\linebreak}

\newenvironment{fminipage}%
  {\begin{Sbox}\begin{minipage}}%
  {\end{minipage}\end{Sbox}\fbox{\TheSbox}}

\newenvironment{algbox}[0]{\vskip 0.2in
\noindent
\begin{fminipage}{6in}
}{
\end{fminipage}
\vskip 0.2in
}

\newenvironment{tightlist}%
{\begin{list}{}{\usecounter{bean}%
  \setlength{\partopsep}{0em}
  \setlength{\leftmargin}{0em}
  \setlength{\labelwidth}{0em}
}}%
{\end{list}}

\newenvironment{tightlist2}%
{\begin{list}{}{\usecounter{bean}%
  \setlength{\topsep}{0em}
  \setlength{\parsep}{0em}
  \setlength{\partopsep}{0em}
  \setlength{\leftmargin}{1.3em}
}}%
{\end{list}}

\newenvironment{tightlistA}%
{\begin{list}{}{\usecounter{bean}%
  \setlength{\topsep}{0em}
  \setlength{\parsep}{0em}
  \setlength{\partopsep}{0em}
  \setlength{\leftmargin}{1.7em}
}}%
{\end{list}}

\newenvironment{tightlistB}%
{\begin{list}{}{\usecounter{bean}%
  \setlength{\topsep}{0em}
  \setlength{\parsep}{0em}
  \setlength{\partopsep}{0em}
}}%
{\end{list}}

{\begin{list}{}{\usecounter{bean}%
  \setlength{\leftmargin}{0.5em}
  \setlength{\labelwidth}{0.5em}
}}%
{\end{list}}

    \newcommand\xx{\boldsymbol{\mathit{x}}}
\newcommand\yy{\boldsymbol{\mathit{y}}}

\newcommand\xxc{\boldsymbol{\mathit{\widetilde{x}}}}
\newcommand\yyc{\boldsymbol{\mathit{\widetilde{y}}}}

\def\mbar{\widehat{m}}

\begin{document}

\title{Lower-Stretch Spanning Trees}

\author{
Michael Elkin\thanks{Part of this work was done in Yale University,
and was supported by the
 DoD University Research Initiative (URI)
administered by the Office of Naval Research under Grant
N00014-01-1-0795. The work was also partially supported by
the Lynn and William Frankel Center for Computer
 Sciences.}\\
Department of Computer Science\\
Ben-Gurion University of the Negev
\and
Yuval Emek\\
Department of Computer Science\\
and Applied Mathematics\\
Weizmann Institute of Science
\and
Daniel A. Spielman\thanks{
Partially supported by NSF grant CCR-0324914.
Part of this work was done at Yale University.}\\
Department of Mathematics \\
Massachusetts Institute of Technology\\
\and
Shang-Hua Teng\thanks{Partially supported by NSF grants CCR-0311430
and ITR CCR-0325630.}\\
Department of Computer Science\\
Boston University and\\
Akamai Technologies Inc.
}

\date{\ }

\maketitle

\begin{abstract}
{
We show that every
  weighted connected graph $G$
  contains as a subgraph a
  spanning tree into which the edges of $G$ can be embedded
  with average stretch
  $O (\log^{2} n \log \log n)$.
Moreover, we show that this tree can be constructed in time
  $O (m \log n + n \log^2 n)$ in general, and in time $O (m\log n)$ if
  the input graph is unweighted.
The main ingredient in our construction is a
   novel graph decomposition technique.

Our new algorithm can be immediately used to
  improve the running time of
  the recent solver for symmetric diagonally dominant linear systems
  of Spielman and Teng from
\[
m 2^{ \left(O (\sqrt{\log n\log\log n}) \right) }\ \mbox{to} \ m \log^{O (1)}n,
\]
and to
  $O (n \log^{2} n \log \log n)$ when the system
  is planar.
Our result can also be used to improve several earlier approximation
  algorithms that use low-stretch spanning trees.
}
\end{abstract}


\vspace{1mm}
\noindent
{\bf Categories and Subject Descriptors:} F.2 {[Theory of Computation]}: {Analysis of Algorithms and
Problem Complexity}

\vspace{1mm}
\noindent
{\bf General Terms:} Algorithms, Theory.

\vspace{1mm}

\section{Introduction}\label{sec:intro}

Let $G = (V,E,w)$ be a weighted connected graph, where $w$ is a function from $E$ into the positive reals. We define the length of each edge $e \in E$ to be the reciprocal of its weight:
\[
d (e) = 1/w (e).
\]

Given a spanning tree $T$ of $V$, we define the distance in $T$
between a pair of vertices $u,v \in V$, $\tdist{T}{u,v}$, to be the
sum of the lengths of the edges on the unique path in $T$ between $u$
and $v$. We can then define the stretch\footnote{Our definition
  of the stretch differs slightly from that used in \cite{AKPW}:
  ${\tdist{T}{u,v}}/{\tdist{G}{u,v}}$, where
  $\tdist{G}{y,v}$ is the length of the shortest-path between $u$ and $v$.
  See Subsection~\ref{subsec:applications} for a discussion of the difference.}
 of an edge $(u,v) \in E$ to be
\[
\stretch{T}{u,v} = \frac{\tdist{T}{u,v}}{d (u,v)},
\]
and the average stretch over all edges of $E$ to be
\[
\avestretch{T}{E} =
  \frac{1}{\sizeof{E}}
  \sum _{(u,v) \in E} \stretch{T}{u,v}.
\]

Alon, Karp, Peleg and West~\cite{AKPW}
  proved that every weighted
  connected graph $G = (V,E,w)$ of $n$ vertices and $m$ edges contains a
spanning tree $T$ such that
\[
\avestretch{T}{E} = \exp \left(O (\sqrt{\log n \log \log n}) \right),
\]
and that there exists a collection $\tau = \{T_1,\ldots,T_h\}$ of
spanning trees of $G$ and a probability distribution $\Pi$ over $\tau$
such that for every edge $e \in E$,
\[
\expec{T  \leftarrow \Pi}{\stretch{T}{e}} = \exp \left(O (\sqrt{\log n
\log \log n}) \right).
\]

The class of weighted graphs considered in this paper
  includes multi-graphs that may contain weighted self-loops
  and multiple weighted-edges between a pair of vertices.
The consideration of multi-graphs is essential
  for several results (including some in \cite{AKPW}).

The result of \cite{AKPW} triggered the study of low-distortion
  embeddings into {\em probabilistic tree metrics}.
Most notable in this context is the work of Bartal \cite{Bartal1, Bartal2}
  which shows that if the requirement that the trees $T$
  be {\em subgraphs} of $G$ is abandoned, then the upper bound of
  \cite{AKPW} can be improved by finding a tree whose distances
  approximate those in the original graph with average distortion
  $O(\log n \cdot \log\log n)$.
On the negative side, a lower bound of
  $\Omega(\log n)$ is known for both scenarios \cite{AKPW,Bartal1}. The
  gap left by Bartal was recently closed  by  Fakcharoenphol, Rao, and
  Talwar \cite{FRT}, who have shown a tight upper bound of $O(\log n)$.

However, some applications of graph-metric-approximation require trees
  that are subgraphs.
Until now, no progress had been made on reducing
  the gap between the upper and lower bounds proved in \cite{AKPW}
  on the average stretch of subgraph spanning trees.
The bound achieved in \cite{AKPW} for general weighted graphs
  had been the best bound known for {\em unweighted} graphs,
  even for {\em unweighted planar} graphs.

In this paper\footnote{
In the submitted version of this paper, we proved the weaker bound
  on average stretch of $O ((\log n \log \log n)^{2})$.
The improvement in this paper comes from re-arranging the arithmetic
  in our analysis.
Bartal~\cite{BartalComm} has obtained a similar improvement by
  other means.
}%
, we significantly narrow this gap by improving the upper
  bound of \cite{AKPW} from $\exp ( O(\sqrt{\log n\log\log n}))$ to
  $O(\log^{2} n \log\log n)$.
Specifically, we give an algorithm
  that for every weighted connected graph $G = (V,E, w)$, constructs a
  spanning tree $T \subseteq E$ that satisfies \AvoidOverfull{}
  \( \avestretch{T}{E} = O( \log^{2} n \log \log n) \).
The running time of our algorithm is
  \( O(m \log n + n\log^2 n) \) for weighted graphs,
  and \( O(m \log n) \) for unweighted.
Note that the input graph need not be simple and its number of edges
   $m$ can be much larger than $n \choose 2$.
However, as  proved in \cite{AKPW}, it is enough to
  consider graphs with at most $n (n+1)$ edges.

We begin by presenting a simpler algorithm that guarantees
  a weaker bound, $\avestretch{T}{E} = O(\log^3 n)$.
As a consequence of the result in \cite{AKPW} that the existence of a
  spanning tree with average stretch $f(n)$ for every weighted graph
  implies the existence of a distribution of spanning trees in which
  every edge has expected stretch $f (n)$,
  our result implies that for every weighted
  connected graph $G = (V,E,w)$ there
  exists a probability distribution $\Pi$ over a set  $\tau =
  \{T_1,\ldots,T_h\}$ of spanning trees ($T \subseteq E$ for every $T
  \in \tau$) such that for every $e \in E$, $\expec{T  \leftarrow \Pi
  }{\stretch{T}{e}} = O(\log^{2} n\log\log n).$
Furthermore, our
  algorithm itself can be adapted to produce a probability distribution
  $\Pi$ that guarantees a slightly weaker bound of $O(\log^3 n)$ in time
  $O(m \cdot \log^2 n)$.
So far, we have not yet been able to verify
  whether our algorithm can be adapted to produce the bound of \AvoidOverfull{}
  $O(\log^{2}  n \cdot \log\log n)$ within similar time limits.

\subsection{Applications} \label{subsec:applications}

For some of the applications listed below it is essential to define
the stretch of an edge \( (u,v) \in E \) as in \cite{AKPW}, namely,
\( \stretch{T}{u,v} = \tdist{T}{u,v} / \tdist{G}{u,v} \). The
algorithms presented in this paper can be adapted to handle this
alternative definition for stretch simply by assigning new weight \(
w'(u,v) = 1 / \tdist{G}{u,v} \) to every edge \( (u,v) \in E \) (the
lengths of the edges remain unchanged). Observe that the new weights
can be computed in a preprocessing stage independently of the
algorithms themselves, but the time required for this computation
may dominant the running time of the algorithms.

\subsubsection{Solving Linear Systems}

Boman and Hendrickson~\cite{BomanHendricksonAKPW}
  were the first to realize that low-stretch spanning trees
  could be used to solve symmetric diagonally dominant linear systems.
They  applied the spanning
  trees of~\cite{AKPW} to design solvers that run in time

\[
m^{3/2} 2^{O (\sqrt{\log n \log \log n})}
  \log (1/\epsilon ),
\]
where $\epsilon $ is the precision of the solution.
Spielman and Teng \cite{SpielmanTengPrecon} improved their results
  to
\[
m 2^{O (\sqrt{\log n\log\log n}) } \log (1/\epsilon).
\]

Unfortunately, the trees produced by the algorithms of Bartal \cite{Bartal1, Bartal2}
   and Fakcharoenphol, Rao, and Talwar \cite{FRT} cannot be used
   to improve these linear solvers, and it is currently not known
   whether it is possible
   to solve linear systems efficiently using trees that are not subgraphs.

By applying the low-stretch spanning trees developed in
  this paper, we can reduce the time for solving these linear systems
  to
\[
m \log^{O (1)}n \log (1/\epsilon),
\]
  and to $O (n \log^{2} n \log \log n \log (1/\epsilon))$ when the
  systems are planar.
Applying a recent reduction of Boman, Hendrickson and Vavasis \cite{BHV},
  one obtains a $O (n \log^{2} n \log \log n  \log (1/\epsilon))$ time algorithm
  for solving the linear systems that arise when
  applying the finite element method to solve two-dimensional elliptic
  partial differential equations.

\subsubsection{Alon-Karp-Peleg-West Game}

Alon, Karp, Peleg and West~\cite{AKPW} constructed low-stretch
spanning trees to upper-bound the value of a zero-sum two-player game
that arose in their analysis of an algorithm for the $k$-server
problem: at each turn, the {\em tree player} chooses a spanning tree
$T$ and the {\em edge player} chooses an edge $e\in E$,
simultaneously. The payoff to the edge player is $0$ if $e \in T$ and
$\stretch{T}{e}+1$ otherwise. They showed that if every
$n$-vertex weighted connected graph $G$  has a spanning tree $T$
of average stretch $f (n)$, then the value of this game is at most \( f
(n) + 1 \). Our new result lowers the bound on the value of this
graph-theoretical game from \( \exp \left(O (\sqrt{\log n \log \log
n}) \right) \) to \( O\left( \log^{2} n\log\log n \right) \).

\subsubsection{MCT Approximation}

Our result can be used to improve drastically the upper bound on the
approximability of the \emph{minimum communication cost spanning tree}
(henceforth, \emph{MCT}) problem. This problem was introduced in
\cite{Hu}, and is listed as [ND7] in \cite{GareyJohnson} and \cite{NPKann}.

The instance of this problem is a weighted graph $G = (V,E,w)$, and a
matrix $\{r(u,v) \mid u,v \in V\}$ of nonnegative requirements. The
goal is to construct a spanning tree $T$ that minimizes $c(T) =
\sum_{u,v \in V} r(u,v) \cdot \tdist{T}{u,v}$.

Peleg and Reshef \cite{PelReshef} developed a
  $2^{O(\sqrt{\log n\cdot \log\log n})}$ approximation
  algorithm for the MCT problem on metrics
   using the result of \cite{AKPW}.
A similar approximation ratio can be achieved for arbitrary graphs.
Therefore our result can be used to produce an
   efficient $O(\log^{2} n  \log\log n )$
  approximation algorithm for the MCT problem on arbitrary graphs.

\subsubsection{Message-Passing Model}

Embeddings into probabilistic tree metrics have been extremely useful
in the context of approximation algorithms
(to mention a few:
buy-at-bulk network design \cite{AwerbuchAzar}, graph Steiner problem
\cite{GargKonjevodRavi}, covering Steiner problem \cite{KonjRavi}).
However, it is not clear that these algorithms
  can be implemented in  the
  message-passing model of distributed computing (see \cite{PelegBook}).
In this model, every vertex of the input graph hosts a processor, and
  the processors communicate over the edges of the graph.

Consequently, in this model executing an algorithm that starts by
  constructing a non-subgraph spanning tree of the network, and then
  solves a problem whose instance is this tree is very problematic,
  since direct communication over the links of this ``virtual'' tree is
  impossible.
This difficulty disappears if the tree in this scheme is a
  {\em subgraph } of the graph.
We believe that our result will
  enable the adaptation of the these approximation algorithms to
  the message-passing model.

\subsection{Our Techniques}

We build our low-stretch spanning trees by recursively applying
  a new graph decomposition
  that we call a \textit{star-decomposition}.
A star-decomposition of a graph is a partition of the vertices
  into sets that are connected into a star: a central set is connected
  to each other set by a single edge (see Figure~\ref{fig:star}).
We show how to find star-decompositions that do not cut too many
short edges and such that the radius of the graph induced by the
star decomposition is not much larger than the radius of the
original graph.

Our algorithm for finding a low-cost star-decomposition applies a
generalization of the ball-growing technique of
Awerbuch~\cite{Awerbuch} to grow \textit{cones}, where the cone at a
vertex $x$ induced by a set of vertices $S$ is the set of vertices
whose shortest path to $S$ goes through $x$.

\subsection{The Structure of the Paper}

In Section \ref{sec:preliminaries}, we define our notation.
In Section \ref{sec:treelogcube}, we introduce
  the star decomposition of a weighted connected graph.
  We then  show how to use this decomposition to
    construct a subgraph spanning tree
    with average stretch $O (\log^{3} n)$.
In Section \ref{sub:cutting}, we present our
  star decomposition algorithm.
In Section \ref{sec:Improved}, we refine our construction and improve
  the average stretch to $O \left(\log^{2} n\log\log n \right)$.
Finally, we conclude the paper in Section
6
and list
  some open questions.

\section{Preliminaries}\label{sec:preliminaries}

Throughout the paper, we assume that the input graph is a weighted
connected multi-graph \( G = (V, E, w) \), where \(w\) is a \emph{weight}
function from \(E\) to the positive reals. Unless stated otherwise, we let
\(n\) and \(m\) denote the number of vertices and the number of edges
in the graph, respectively. The \emph{length} of an edge \( e \in E \) is
defined as the reciprocal of its weight, denoted by \( d(e) = 1 / w(e)
\).

\begin{tightlist}
\item For two vertices $u,v \in V$,
  we define $\dist{u}{v}$ to be the length of the shortest
  path between $u$ and $v$ in $E$.
  We write $\tdist{G}{u,v}$ to emphasize that the distance is in the
  graph $G$.

\item
For a set of vertices, $S \subseteq V$,
  $G (S)$ is the subgraph induced by vertices in $S$.
  We write $\tdist{S}{u,v}$ instead of $\tdist{G (S)}{u,v}$
  when $G$ is understood.
  $E (S)$ is the set of edges with both endpoints in $S$.

\item
For $S, T \subseteq V$,
$E (S,T)$ is the set of edges with one endpoint in $S$
  and the other in $T$.

\item The \emph{boundary} of a set $S$, denoted $\bdry{S}$, is the set of edges
with exactly one endpoint in $S$.

\item
For a vertex $x\in V$, $\tdist{G}{x,S}$ is the length of the shortest path
  between $x$ and a vertex in $S$.

\item A \emph{multiway partition} of $V$ is a collection
  of pairwise-disjoint sets
  $\setof{\vs{V}{1}{k}}$ such that
  $\bigcup_{i} V_{i}= V$.
The \emph{boundary} of a multiway partition, denoted
  $\bdry{\vs{V}{1}{k}}$, is the set of edges with endpoints
  in different sets in the partition.

\item
The \emph{volume} of a set \(F\) of edges, denoted $\vol{F}$, is the
size of the set $|F|$.

\item
The \emph{cost} of a set \(F\) of edges, denoted \( \cost{F} \), is
the sum of the weights of the edges in \(F\).

\item
The \emph{volume} of a set \(S\) of vertices, denoted \( \vol{S} \),
is the number of edges with at least one endpoint in \(S\).

\item
The \emph{ball} of radius $r$ around a vertex $v$, denoted $B
(r,v)$, is the set of vertices of distance at most $r$ from $v$.

\item
The \emph{ball shell} of radius \(r\) around a vertex \(v\), denoted
\( \BallShell{r}{v} \), is the set of vertices right outside \( B(r,v) \),
that is, \( \BallShell{r}{v} \) consists of every vertex \( u \in V - B(r,v)
\) with a neighbor \( w \in B(r,v) \) such that \( \dist{v}{u} =
\dist{v}{w} + d(u,w) \).

\item
For $v\in V$, $\Gradius{G}{v}$ is the smallest $r$ such that every
vertex of $G$ is within distance $r$ from $v$. For a set of vertices
\( S \subseteq V \), we write \( \Gradius{S}{v} \) instead of \(
\Gradius{G (S)}{v} \) when \(G\) is understood.
\end{tightlist}

\section{Spanning-Trees of \mbox{${\bf O ({\bf
\lowercase{log}}^{3}\lowercase{n})}$} Stretch}\label{sec:treelogcube}

We present our first algorithm that
  generates a spanning tree with average stretch $O
  (\log^{3}n)$.
We first state the properties of the graph decomposition algorithm
  at the heart of our construction.
We then present the construction and analysis of the
  low-stretch spanning trees.
We defer the description of the graph decomposition algorithm
  and its analysis
  to Section~\ref{sub:cutting}.

\subsection{Low-Cost Star-Decomposition}\label{sub:structure}

\begin{figure}[htop]
\centering\epsfig{figure=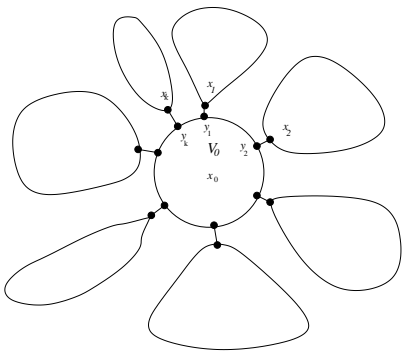,height=1.8in}
\caption{Star Decomposition. \label{fig:star}}
\end{figure}

\begin{definition}[Star-Decomposition]
A multiway partition
$\{\vs{V}{0}{k}\}$ is a {\em star-decomposition} of a weighted
  connected graph $G$ with center
  $x_{0} \in V$
  (see Figure \ref{fig:star}) if $x_{0} \in V_{0}$ and
\begin{itemize}
\item [1.] for all $0\leq i\leq k$, the subgraph induced on $V_{i}$ is
   connected, and
\item [2.] for all $i\geq  1$, $V_{i}$ contains an \texttt{anchor}
vertex $x_{i}$ that is connected to a vertex $y_{i} \in V_{0}$ by an
edge
  $(x_{i},y_{i}) \in E$.
  We call the edge $(x_{i}, y_{i})$ the \texttt{bridge} between
  $V_{0}$ and $V_{i}$.
\end{itemize}
Let $r = \Gradius{G}{x_0}$, and $r_i = \Gradius{V_i}{x_i}$ for each
  $0 \le i \le k$.
For $\delta ,\epsilon \leq 1/2$, a star-decomposition
$\{\vs{V}{0}{k}\}$ is a $(\delta,\epsilon )$-star-decomposition if
\begin{itemize}
\item [a.]  $\delta r\leq r_{0}\leq (1-\delta ) r$, and

\item [b.]
$r_{0}+ d (x_{i},y_{i}) +  r_{i} \leq (1+\epsilon ) r$, for each $i \geq 1$.
\end{itemize}
The \texttt{cost} of the star-decomposition is
   $\cost{\bdry{\vs{V}{0}{k}}}$.
\end{definition}

Note that if
 $\setof{\vs{V}{0}{k}}$ is a $(\delta ,\epsilon )$ star-decomposition
  of $G$, then the graph consisting of the union of the induced subgraphs on
  $\vs{V}{0}{k}$ and the bridge edges $(y_{k}, x_{k})$ has radius at most
  $(1+\epsilon)$ times the radius of the original graph.

In Section~\ref{sub:cutting}, we present an algorithm
  \texttt{StarDecomp} that satisfies the following
  cost guarantee.
Let $\xx= (\vs{x}{1}{k})$ and $\yy = (\vs{y}{1}{k})$.

\begin{lemma}[Low-Cost Star Decomposition]\label{lem:stardecomposition}
Let $G = (V,E,w)$ be a connected weighted graph and let \(x_0\) be a
vertex in \(V\). Then for every positive \( \epsilon \leq 1 / 2 \),
\[
  \left(\setof{\vs{V}{0}{k}},\xx,\yy\right)
  = \mathtt{starDecomp} (G, x_{0}, 1/3, \epsilon) ~ ,
\]
in time $O (m + n \log n)$, returns a  $(1/3, \epsilon)$-star-decomposition of
  $G$ with center $x_0$ of cost
\[
\cost{\bdry{V_0, \ldots, V_k}} ~ \leq ~ \frac{6 \, m \log_{2}
(m+1)}{\epsilon \cdot \Gradius{G}{x_{0}}}.
\]
On unweighted graphs, the running time is $O(m)$.
\end{lemma}

\subsection{A Divide-and-Conquer Algorithm}\label{sub:log3}

The basic idea of our algorithm is to use low-cost
star-decomposition in a divide-and-conquer (recursive) algorithm to
construct a spanning tree. We use \( \widehat{n} \) (respectively, \(
\widehat{m} \)) rather than \(n\) (resp., \(m\)) to distinguish the
number of vertices (resp., number of edges) in the original graph,
input to the first recursive invocation, from that of the graph
input to the current one.

Before invoking our algorithm we apply a linear-time transformation
from \cite{AKPW} that transforms the graph into one with at most \(
\widehat{n} (\widehat{n} + 1) \) edges (recall that a multi-graph of
\(\widehat{n}\) vertices may have an arbitrary number of edges), and
such that the average-stretch of the spanning tree on the original
graph will be at most twice the average-stretch on this graph.

We begin by showing how to construct low-stretch spanning trees
  in the case that all edges have length $1$.
In particular, we use the fact that in this case
  the cost of a set of edges equals the number of edges in the set.

\begin{algbox}
\noindent Fix $\alpha = (2 \log_{4/3} (\widehat{n}+6))^{-1}$.
\vskip 0.1in
\noindent \( T = \mathtt{UnweightedLowStretchTree} (G, x_{0}) \).
\vskip 0.1in
\begin{tightlistA}
\item [0.] If $\sizeof{V} \leq 2$, return $G$. (If \(G\) contains
multiple edges, return a single copy.)
\item [1.] Set $\rho = \Gradius{G}{x_{0}}$.
\item [2.]
$(\setof{\vs{V}{0}{k}},\xx,\yy)$~$=\mathtt{StarDecomp}(G, x_{0}, 1/3, \alpha)$
\item [3.]
  For $0 \leq i \leq k$, \\
  set  $T_{i} = \mathtt{UnweightedLowStretchTree} (G (V_{i}), x_{i})$

\item [4.] Set $T = \Union_{i} T_{i} \union \Union_{i} (y_{i},x_{i})$.
\end{tightlistA}
\end{algbox}

\begin{theorem}[Unweighted]\label{thm:Unweighted}
Let \( G = (V, E) \) be an unweighted connected graph and let \(x_0\) be a
vertex in \(V\). Then
\[
T = \mathtt{UnweightedLowStretchTree} (G, x_{0}) ~ ,
\]
in time $O (\widehat{m} \log \widehat{n})$, returns a spanning tree of
\(G\) satisfying
\begin{equation} \label{eq:RadiusUnweighted}
\Gradius{T}{x_{0}} \leq \sqrt{e} \cdot \Gradius{G}{x_{0}}
\end{equation}
and
\begin{equation} \label{eq:AverageStretchUnweighted}
\avestretch{T}{E} \leq O (\log^{3} \widehat{m}) ~ .
\end{equation}
\end{theorem}
\begin{proof}
For our analysis, we define a family of graphs that converges to $T$.
For a graph $G$, we let
\[
 (\setof{\vs{V}{0}{k}}, \xx, \yy  )
  = \mathtt{StarDecomp} (G, x_{0}, 1/3, \alpha)
\]
and recursively define
\[
R_{0} (G) ~=~ G \quad \mbox{and}\quad ~~
R_{t} (G) ~= ~  \Union_{i} (y_{i},x_{i}) \union \Union_{i} R_{t-1} (G
(V_{i})) ~ .
\]
The graph~$R_{t} (G)$ is what one would obtain if we forced
  \texttt{UnweightedLowStretchTree} to return its input graph after
  $t$ levels of recursion.

Because for all $\widehat{n}\geq 0$, $(2\log_{4/3}
(\widehat{n}+6))^{-1}\leq 1/12$, we have $(2/3+\alpha )\leq 3/4$.
Thus, the depth of the recursion in \texttt{UnweightedLowStretchTree}
  is at most $\log_{4/3} \widehat{n}$, and we have
  $R_{\log_{4/3} \widehat{n}} (G) = T$.
One can prove by induction that, for every \( t \geq 0 \),
\[
  \Gradius{  R_{t} (G)}{x_{0}}
  \leq (1+\alpha)^{t} \Gradius{G}{x_{0}}.
\]
The claim in \eqref{eq:RadiusUnweighted} now follows from $(1+\alpha
)^{\log_{4/3} \widehat{n}} \leq  \sqrt{e}$.
To prove the claim in \eqref{eq:AverageStretchUnweighted}, we note that
\begin{eqnarray}
\label{eq:str1}
\sum_{(u,v) \in \bdry{\vs{V}{0}{k}}} \stretch{T}{u,v} & \leq &
\sum_{(u,v) \in \bdry{\vs{V}{0}{k}}} \left( \tdist{T}{x_{0},u} +
\tdist{T}{x_{0},v} \right) \\
\label{eq:str3}
& \leq  &
  \sum_{(u,v) \in \bdry{\vs{V}{0}{k}}}
   2 \sqrt{e} \cdot \Gradius{G}{x_{0}},
  \text{ by \eqref{eq:RadiusUnweighted}} \\
& \leq &
   2 \sqrt{e} \cdot  \Gradius{G}{x_0}
\left(  \frac{
      6 \, m \log_{2} (\widehat{m}+1)
       }{
        \alpha \cdot \Gradius{G}{x_0}
       }
 \right),
  \text{ by Lemma~\ref{lem:stardecomposition}. } \nonumber
\end{eqnarray}
Applying this inequality to all graphs at all $\log_{4/3} \widehat{n}$
levels of the recursion, we obtain
\[
  \sum_{(u,v) \in E}
  \stretch{T}{u,v}
~ \leq~
24 \, \sqrt{e} \, \widehat{m} \log_{2} \widehat{m} \log_{4/3} \widehat{n}
\log_{4/3} (\widehat{n}+6)
~ =~
   O ( \widehat{m} \log^{3} \widehat{m}) ~ .
\]
\end{proof}

We now extend our algorithm and proof to general weighted connected
   graphs.
We begin by pointing out a subtle difference between general
   and unit-weight graphs.
In our analysis of
  \texttt{UnweightedLowStretchTree}, we used the facts that
  $\Gradius{G}{x_{0}} \leq n$ and that each edge length is
  1 to show that the depth of recursion is at most $\log_{4/3}n$.
In general weighted graphs, the ratio of $\Gradius{G}{x_{0}}$
  to the length of the shortest edge can be arbitrarily large.
Thus, the recursion can be very deep.
To compensate, we will contract all edges that are significantly
  shorter than the radius of their component.
In this way, we will guarantee that each edge is only active
  in a logarithmic number of iterations.

Let $e = (u,v)$ be an edge in $G= (V,E,w)$.
The contraction of $e$ results in a new graph
  by identifying $u$ and $v$ as a new vertex
  whose neighbors are the union of the neighbors of $u$ and $v$.
All self-loops created by the contraction are discarded.
We refer to  $u$ and $v$ as the preimage of
  the new vertex.

We now state and analyze our algorithm for
  general weighted graphs.

\begin{algbox}
\noindent Fix \( \beta = (2 \log_{4/3} (\widehat{n} + 32))^{-1} \).
\vskip 0.1in \noindent \( T = \mathtt{LowStretchTree} (G = (V, E,w),
x_{0}) \).
\vskip 0.1in
\begin{tightlistA}
\item [0.] If $\sizeof{V} \leq 2$, return $G$. (If \(G\) contains
multiple edges, return the shortest copy.)
\item [1.] Set $\rho = \Gradius{G}{x_{0}}$.
\item [2.] Let \( \Contracted{G} = (\Contracted{V}, \Contracted{E}) \)
  be the graph obtained by contracting
  all edges in $G$ of length less than \( \beta \rho / \widehat{n} \).
\item [3.]
  \( \left(\setof{\vs{\Contracted{V}}{0}{k}},
  \xxc  , \yyc \right) =
  \mathtt{StarDecomp} (\Contracted{G}, x_{0}, 1/3, \beta) \).
\item [4.] For each $i$, let $V_{i}$ be the preimage
  under the contraction of step 2
  of vertices in $\Contracted{V}_{i}$, and
  $(x_{i},y_{i}) \in V_{0} \times V_{i}$ be the edge of shortest
  length for which $x_{i}$ is a preimage of $\Contracted{x}_{i}$ and
  $y_{i}$ is a preimage of $\Contracted{y}_{i}$.
\item [5.]
  For $0 \leq i \leq k$,
set \( T_{i} = \mathtt{LowStretchTree} (G
  (V_{i}), x_{i}) \)
\item [6.] Set $T = \Union_{i} T_{i} \union \Union_{i} (y_{i},x_{i})$.
\end{tightlistA}
\end{algbox}

In what follows, we refer to the graph \( \Contracted{G} \),
obtained by contracting some of the edges of the graph \(G\), as the
\emph{edge-contracted graph}.

\begin{theorem}[Low-Stretch Spanning Tree]\label{thm:main}
Let  $G= (V,E,w)$ be a weighted connected graph and let \( x_0 \) be a
vertex in \(V\). Then
\[
T = \mathtt{LowStretchTree}(G,x_{0}) ~ ,
\]
in time $O (\widehat{m} \log \widehat{n} + \widehat{n} \log^2
\widehat{n})$, returns a spanning tree of $G$ satisfying
\begin{equation} \label{eq:RadiusWeighted}
\Gradius{T}{x_0} \leq 2 \sqrt{e} \cdot  \Gradius{G}{x_0}
\end{equation}
and
\begin{equation} \label{eq:AverageStretchWeighted}
\avestretch{T}{E} = O (\log^{3} \widehat{m})
\end{equation}
\end{theorem}


\begin{proof}

We first establish notation similar to that used in the proof of
Theorem~\ref{thm:Unweighted}.
Our first step is to define a procedure, \texttt{SD}, that captures
  the action of the algorithm in steps 2 through 4.
We then define $R_{0} (G) = G$ and
\[
  R_{t} (G) =   \Union_{i} (y_{i},x_{i}) \union
                \Union_{i} R_{t-1} (G (V_{i})),
\]
where
\[
 (\setof{\vs{V}{0}{k}}, \vs{x}{1}{k}, \vs{y}{1}{k})
  = \mathtt{SD} (G, x_{0}, 1/3, \beta).
\]

We now prove the claim in \eqref{eq:RadiusWeighted}.
Let \( \rho = \Gradius{G}{x_0} \). Let \( t = 2
\log_{4/3}{(\widehat{n} + 32)} \) and let \( \rho_{t} = \Gradius{R_{t}
(G)}{x_{0}} \). Each contracted edge is of length at most \( \beta
\rho / \widehat{n} \), and every path in the graph \( G(V_i) \)
contains at most \(n\) contracted edges, hence the insertion of the
contracted edges to \( G(V_i) \) increases its radius by an additive factor of at most \( \beta \rho \). Since \( (2 \log_{4/3} (\widehat{n} + 32))^{-1}
\leq 1 / 24 \) for every \( n \geq 0 \), it follows that \( 2 / 3 + 2
\, \beta \leq 3 / 4 \). Therefore, following the proof of
Theorem~\ref{thm:Unweighted}, we can show that \( \rho_t \) is at most \(
\sqrt{e} \cdot \Gradius{G}{x_{0}} \).

We know that each component of \(G\) that remains after \(t\) levels
of the recursion has radius at most \( \rho (3 / 4)^t \leq \rho
/ \widehat{n}^{2} \). We may also assume by induction that for the
graph induced on each of these components, \texttt{LowStretchTree}
outputs a tree of radius at most \( 2 \sqrt{e} (\rho /
\widehat{n}^{2}) \). As there are at most \(n\) of these components,
we know that the tree returned by the algorithm has radius at most
\[
\sqrt{e} \, \rho + n \times 2 \sqrt{e} (\rho / \widehat{n}^{2}) ~ \leq ~ 2
\, \sqrt{e} \, \rho ~ ,
\]
for $\widehat{n} \geq 2$.

We now turn to the claim in \eqref{eq:AverageStretchWeighted}, the
bound on the stretch.
In this part, we let \( E_{t} \subseteq E \) denote the set of edges
  that are present at recursion depth \(t\).
That is, their endpoints are
  not identified by the contraction of short edges in step~2, and their
  endpoints remain in the same component.
We now observe that no edge can be present at more than \( \log_{4 /
3}{((2 \, \widehat{n} / \beta) + 1)} \) recursion depths. To see this,
consider an edge \( (u, v) \) and let \(t\) be the first recursion
level for which the edge is in \( E_{t} \). Let \( \rho_t \) be the
radius of the component in which the edge lies at that time. As \(u\)
and \(v\) are not identified under contraction, they are at distance
at least \( \beta \rho_t / \widehat{n} \) from each other. (This
argument can be easily verified, although the condition for edge
contraction depends on the length of the edge rather than on the
distance between its endpoints.) If \(u\) and \(v\) are still in the same graph on recursion level \( t + \log_{4 / 3}{((2 \, \widehat{n} / \beta) +
1)} \), then the radius of this graph is at most \( \rho_t / ((2 \,
\widehat{n} / \beta) + 1) \), thus its diameter is strictly less than
\( \beta \rho_t / \widehat{n} \), in contradiction to the distance
between \(u\) and \(v\).

Similarly to the way that $ \sum_{(u,v) \in \bdry{\vs{V}{0}{k}}}
  \stretch{T}{u,v}$ is upper-bounded in inequalities \eqref{eq:str1}-
\eqref{eq:str3}
 in the proof of Theorem~\ref{thm:Unweighted}, it follows  that
  the total contribution to the stretch at depth $t$
  is at most
\[
O (\vol{E_{t}} \log^{2}\widehat{m}).
\]
Thus, the sum of the stretches over all recursion depths
  is
\[
  \sum _{t} O \left( \vol{E_{t}} \log^{2}\widehat{m} \right)
  = O\left(\widehat{m} \log^{3}\widehat{m}\right).
\]

We now analyze the running time of the algorithm. On each recursion
level, the dominant cost is that of performing the \texttt{StarDecomp}
operations on each edge-contracted graph. Let \(V_t\) denote the
set of vertices in all edge-contracted graphs on recursion level
\(t\). Then the total cost of the \texttt{StarDecomp} operations on
recursion level $t$ is at most $O(\sizeof{E_{t}} + \sizeof{V_t} \log
\sizeof{V_t})$. We will prove soon that \( \sum_{t}{\sizeof{V_t}} = O
(\widehat{n} \log{\widehat{n}}) \), and as $\sum_{t} \sizeof{E_{t}} = O
(\widehat{m} \log \widehat{m})$, it follows that the total running
time is $O (\widehat{m} \log \widehat{n} + \widehat{n} \log^2 \widehat{n})$.
Note that for unweighted graphs $G$, the running time is only
$O(\widehat{m} \log \widehat{n})$.
\end{proof}


The following lemma shows that even though the number of recursion
levels can be very large, the overall number of vertices in
edge-contracted graphs appearing on different recursion levels is at
most $O(\widehat{n} \log \widehat{n})$. This lemma is used only for
the analysis of the running time of our algorithm; a reader interested
only in the existential bound can skip it.

\begin{lemma}
\label{lm:num_vs}
Let \( V_t \) be the set of vertices appearing in edge-contracted
graphs on recursion level $t$. Then $\sum_t \sizeof{V_t} =
O(\widehat{n} \log \widehat{n})$.
\end{lemma}
\begin{proof}
Consider an edge-contracted graph \( \Contracted{G} = (\Contracted{V},
\Contracted{E}) \) on recursion level \(t\) and let \(x\) be a vertex
in \( \Contracted{V} \). The vertex \(x\) was formed as a result of a
number of edge contractions (maybe \(0\)). Consequently, \(x\) can be
viewed as a set of all original vertices that were identified together
to form it, i.e., \(x\) can be viewed as a \emph{super-vertex}. Let \(
\OriginalVertices{x} \) denote the set of original vertices that were
identified together to form the super-vertex \( v \in \Contracted{V}
\).

We claim that for every super-vertex \( x \in V_{t + 1} \), there
exist a super-vertex \( y \in V_t \) such that \(
\OriginalVertices{x} \subseteq \OriginalVertices{y} \). To prove it,
note that every graph on recursion level \(t + 1\) corresponds to a
single component of a star decomposition on recursion level \(t\).
Moreover, an edge that was contracted on recursion level \(t + 1\)
must have been contracted on recursion level \(t\) as well.
Therefore we can consider a directed forest \( \mathcal{F} \) in
which every node at depth \(t\), corresponds to some super-vertex in
\( V_t \), and an edge leads from a node \(y\) at depth \(t\) to a
node \(x\) at depth \(t + 1\), if \( \OriginalVertices{x} \subseteq
\OriginalVertices{y} \). Note that the roots of \( \mathcal{F} \)
correspond to super-vertices on recursion level \(0\) and the leaves
of \( \mathcal{F} \) correspond to the original vertices of the
graph \(G\).

In the proof of Theorem~\ref{thm:main}, we showed that every edge is
present on \( O (\log{\widehat{n}}) \) recursion levels. Following a
similar line of arguments, one can show that every super-vertex is
present on \( O (\log{\widehat{n}}) \) (before it decomposes to
smaller super-vertices, each contains a subset of its vertices). Since
there are \( \widehat{n} \) vertices in the original graph \(G\),
there are \( \widehat{n} \) leaves in \( \mathcal{F} \). Therefore the
overall number of nodes in the directed forest \( \mathcal{F} \) is \(
O (\widehat{n} \log{\widehat{n}}) \), and the bound on \(
\sum_{t}{\sizeof{V_t}} \) holds.

%
%
\end{proof}

\section{Star decomposition}\label{sub:cutting}

Our star decomposition algorithm exploits two algorithms for growing
  sets.
The first, \texttt{BallCut}, is the standard ball growing technique
  introduced by Awerbuch~\cite{Awerbuch}, and was the basis
  of the algorithm of~\cite{AKPW}.
The second, \texttt{ConeCut}, is a generalization of ball growing
  to cones.
So that we can analyze this second algorithm, we abstract the analysis
  of ball growing from the works of~\cite{LeightonRao,AumannRabani,LLR}.
Instead of nested balls, we consider \textit{concentric systems},
  which we now define.

\begin{definition}[Concentric System]\label{def:concentric}
A {\em concentric system } in a weighted graph $G= (V,E,w)$ is a
family of vertex sets $\calL =\{L_{r} \subseteq V: r \in
\Reals{+}\cup \{0 \} \}$ such that
\begin{enumerate}
\item \( L_0 \neq \emptyset \),
\item $L_{r} \subseteq L_{r'}$ for all $r< r'$, and
\item if a vertex $u \in L_{r}$ and $(u,v)$ is an edge in $E$ then $v
\in L_{r+d(u,v)}$.
\end{enumerate}
\end{definition}

For example, for any vertex $x\in V$, the set of balls
  $\setof{B (r,x)}$ is a concentric system.
The {\em radius} of a concentric system $\calL $ is
   $radius (\calL ) = \min \{r: L_{r} = V \}$.
For each vertex $v\in V$, we define $\gennorm{\calL}{v}$
  to  be the smallest $r$ such that
  $v\in L_{r}$.

\begin{lemma}[Concentric System Cutting]\label{lem:cuttingconcentricsystem}
Let $G = (V,E,w)$ be a connected weighted graph and let $\calL  =
\{L_{r} \}$ be a concentric system. For every two reals \( 0 \leq
\lambda < \lambda' \), there exists a real \( r \in [\lambda,
\lambda') \) such that
\[
\cost{\bdry{L_{r}}}
~ \leq ~
\frac{\vol{L_{r}} + \tau }{\lambda' - \lambda}
\max \left[1,
 \log_{2} \left(\frac{m+\tau }{\vol{E (L_{\lambda})} +\tau } \right) \right]
 ,
\]
where $m = \sizeof{E}$ and
\[
  \tau  = \begin{cases}
1 & \text{if $\vol{E (L_{\lambda})} = 0$}\\
0 & \text{otherwise.}
\end{cases}
\]
\end{lemma}

\begin{proof}
Note that rescaling terms does not effect the statement of the
  lemma.
For example, if all the weights are doubled, then the costs are
  doubled but the distances are halved.
Moreover, we may assume that \( \lambda' \leq radius (\calL) \), since
otherwise, choosing \( r = radius (\calL) \) implies that \(
\bdry{L_r} = \emptyset \) and the claim holds trivially.

Let $r_{i} = \norm{v_{i}}$, and assume that the vertices are ordered
  so that
  $r_{1} \leq r_{2} \leq \dotsb \leq r_{n}$.
We may now assume without loss of generality that each edge
  in the graph has minimal length.
That is, an edge from vertex $i$ to vertex $j$ has length
  $\abs{r_{i} - r_{j}}$.
The reason we may make this assumption is that it only increases
  the weights of edges, making our lemma strictly more difficult to
  prove.
(Recall that the weight of an edge is the reciprocal of its length.)

Let $B_{i} = L_{r_{i}}$.
Our proof will make critical use of a quantity $\mu_{i}$, which is defined
  to be
\[
\mu_i ~=~
%
\tau  +
 \vol{E(B_i)}
 +
  \sum_{(v_{j},v_{k}) \in E: j \leq  i < k}
   \frac{r_{i} - r_{j}}{r_{k} - r_{j}}.
\]
That is, $\mu_{i}$ sums the edges inside $B_{i}$, proportionally counting
  edges that are split by the boundary of the ball.
The two properties of $\mu_{i}$ that we exploit are
\begin{equation}\label{eqn:muiPlus}
  \mu_{i+1} = \mu_{i} + \cost{\bdry{B_{i}}} (r_{i+1} - r_{i}),
\end{equation}
and
\begin{equation}\label{eqn:muiVol}
\tau +  \vol{E (B_{i})} \leq   \mu_{i} \leq \tau  + \vol{B_{i}}.
\end{equation}
The equality~\eqref{eqn:muiPlus}  follows  from the definition
by a straight-forward calculation, as

\[
\mu_i ~= ~ \tau
  + \vol{ E (B_{i+1}) } - \vol{ \{ (v_j, v_{i+1}) \in E \mid j \leq i \} }
   + \sum_{(v_j, v_k) \in E : j \leq i < k}{\frac{r_i - r_j}{r_k - r_j}}
\]
and
\[
\cost{\bdry{B_{i}}} (r_{i+1} - r_{i}) = \sum_{(v_j, v_k) \in E : j
\leq i < k}{\frac{r_{i+1} - r_i}{r_k - r_j}} ~ .
\]



Choose $a$ and $b$ so that \( r_{a-1} \leq \lambda < r_{a} \) and \(
r_{b} < \lambda' \leq r_{b+1} \). Let \( \nu = \lambda' - \lambda \).
We first consider the trivial case in which $b < a$.
In that case, there is no vertex whose distance from $v_{0}$
  is between \(\lambda\) and \(\lambda'\).
Thus every edge crossing \( L_{(\lambda + \lambda') / 2} \)
  has length at least \(\nu\), and therefore cost at most \( 1 / \nu \).
Therefore, by setting \( r = (\lambda + \lambda') / 2 \), we obtain
\[
\cost{\bdry{L_{r}}} \leq \vol{\bdry{L_{r}}} \frac{1}{\nu} \leq
\vol{L_{r}} \frac{1}{\nu} ~ ,
\]
establishing the lemma in this case.

We now define
\[
\eta = \log _{2} \left(\frac{m+\tau }{\vol{E (B_{a-1})} + \tau } \right) ~ .
\]
Note that \( B_{a - 1} = L_{\lambda} \), by the choice of \(a\).
A similarly trivial case is when $[a,b]$ is non-empty, and where
  there exists an $i \in [a - 1, b]$ such that
\[
r_{i + 1} - r_{i} \geq \frac{\nu}{\eta} ~ .
\]
In this case, every edge in $\bdry{B_i}$ has cost at most \( \eta
/ \nu \), and by choosing $r$ to be \( \max\{r_i, \lambda\} \), we
satisfy
\[
  \cost{\bdry{L_{r}}} \
\leq
  \sizeof{\bdry{L_{r}}} \frac{\eta}{\nu}
\leq
  \vol{L_{r}} \frac{\eta}{\nu} ~ ,
\]
hence, the lemma is established in this case.

In the remaining case that
  the set $[a,b]$ is non-empty
  and for all $i \in [a - 1, b]$,
\begin{equation}\label{eqn:shortDist}
  r_{i + 1} - r_{i} < \frac{\nu}{\eta} ~ ,
\end{equation}
  we will prove that there exists an $i \in [a - 1, b]$ such that
\[
  \cost{\bdry{B_{i}}} \leq \frac{\mu_{i} \eta}{\nu} ~ ,
\]
hence, by choosing \( r = \max\{r_i, \lambda\} \), the lemma is
established due to \eqref{eqn:muiVol}.

Assume by way of contradiction
  that
\[
  \cost{\bdry{B_{i}}} > \mu_{i} \eta / \nu
\]
for all $i \in [a - 1, b]$.
It follows, by \eqref{eqn:muiPlus}, that
\[
  \mu_{i + 1} > \mu_{i} + \mu_{i} (r_{i + 1} - r_{i}) \eta  / \nu
\]
for all \( i \in [a - 1, b] \), which implies
\begin{align*}
\mu_{b+1} & > ~ \mu_{a-1} \prod_{i = a - 1}^{b} \left( 1 + (r_{i + 1} - r_{i})
\eta / \nu \right) \\
& \geq ~ \mu_{a-1} \prod_{i = a - 1}^{b} 2^{\left((r_{i + 1} - r_{i})
\eta / \nu \right)} \text{ , by~\eqref{eqn:shortDist} and since \( 1 +
x \geq 2^x \) for every \( 0 \leq x \leq 1 \)} \\
& = ~ \mu_{a-1} \cdot 2^{(r_{b+1} - r_{a-1}) \eta / \nu} \\
& \geq ~ \mu_{a-1} \cdot \left( (m+\tau ) / (\vol{E (B_{a-1})}+\tau )
\right) \\
& \geq ~ m +\tau  \text{ , by \eqref{eqn:muiVol},}
\end{align*}
which is a contradiction.
\end{proof}

An analysis of the following standard ball growing algorithm
  follows immediately by applying
  Lemma~\ref{lem:cuttingconcentricsystem} to the concentric system
$\setof{B (r,x)}$.

\begin{algbox}\label{alg:ballcut}
\( r = \mathtt{BallCut} (G, x_{0}, \rho, \delta) \)
\vskip 0.1in
\begin{tightlist2}
\item [1.] Set \( r = \delta \rho \).
\item [2.] While $\cost{\bdry{B (r, x_{0})}} >
\frac{\vol{B (r,x_{0})} +1}{(1 - 2 \, \delta) \rho }
 \log_{2} (m+1)$,
\begin{tightlistA}
\item [a.] Find the vertex $v \not \in B (r, x_{0})$ that minimizes
$\dist{x_{0}}{v}$ and set $r = \dist{x_{0}}{v}$.
\end{tightlistA}
\end{tightlist2}
\end{algbox}

\begin{corollary}[Weighted Ball Cutting]\label{cor:weightedBallCutting}
Let $G = (V,E,w)$ be a connected weighted graph, let $x\in V$, \(
\rho = \Gradius{G}{x} \), \( r = \mathtt{BallCut} (G, x_{0},
\rho, 1 / 3) \), and $V_{0} = B (r, x)$.
Then \( \rho / 3 \leq r < 2 \, \rho / 3 \) and
\[
\cost{\bdry{V_{0}}} ~ \leq ~ \frac{3 \, (\vol{V_{0}}+1)
  \log_{2} (\sizeof{E} + 1)} {\rho} ~ .
\]
\end{corollary}

We now examine the concentric system that enables us to
  construct $\vs{V}{1}{k}$ in Lemma \ref{lem:stardecomposition}.
\begin{definition}[Ideals and Cones]\label{def:C}
For any weighted graph $G = (V,E,w)$ and $S\subseteq V$,
  the set of {\em forward edges induced by $S$} is
\[
F(S) = \{(u\rightarrow v): (u,v)\in E, \dist{u}{S} + d(u,v) = \dist{v}{S}\}.
\]
For a vertex \( v \in V \), the \emph{ideal of $v$ induced by $S$},
  denoted \( I_{S} (v) \), is the set of vertices reachable from $v$ by
  directed edges in $F(S)$, including $v$ itself.

For a vertex \( v \in V \), the {\em cone of width $l$ around $v$
  induced by $S$}, denoted \( C_{S} (l,v) \),
  is the set of vertices in $V$ that
  can be reached from $v$ by a path, the sum of the lengths of whose
  edges $e$ that do not belong to $F(S)$ is at most $l$.
Clearly, $C_{S} (0,v) = I_{S} (v)$ for all $v \in V$.
\end{definition}
That is, $I_{S} (v)$ is the set of vertices that have shortest
  paths to $S$ that intersect $v$.
Also, $u \in C_{S} (l,v)$ if there exist
  $a_{0}, \dots , a_{k-1}$ and $b_{1}, \dots , b_{k}$
  such that $a_{0} = v$,
  $b_{k} = u$,
  $b_{i+1} \in I_{s} (a_{i})$, $(b_{i}, a_{i}) \in E$,
  and
\[
  \sum_{i} d(b_{i}, a_{i}) \leq l.
\]

We now establish that these cones form concentric systems.

\begin{proposition}[Cones are concentric]\label{pro:concentriccone}
Let $G = (V,E,w)$ be a weighted graph and let $S \subseteq V$.
Then for all $v \in V$, $\setof{C_{S} (l,v)}_{l}$ is a concentric
system in $G$.
\end{proposition}
\begin{proof}
Clearly, $C_{S} (l,v) \subseteq C_{S} (l',v)$ if $l<l'$.
Moreover, suppose $u\in C_{S} (l,v)$ and $(u,w) \in E$.
Then if $(u \rightarrow w) \in F$, then $w \in C_{S} (l,v)$ as well.
Otherwise, the path witnessing that $u \in C_{S} (l,v)$ followed by
 the edge $(u,w)$ to $w$ is a witness that $w\in C_{S} (l+d(u,w),v)$.
\end{proof}
\begin{algbox}\label{alg:conecut}
\( r = \mathtt{ConeCut} (G, v, \lambda, \lambda', S) \)
\vskip 0.1in
\begin{tightlist2}
\item [1.] Set \( r = \lambda \) and
if $\vol{E (C_{S} (\lambda ,v))} = 0$,
\begin{tightlistA}
\item Set $\mu =
 (\vol{C_{S} (r, v)}+1) \log_{2} (m+1)$
\end{tightlistA}
otherwise,
\begin{tightlistA}
\item Set $\mu =
\vol{C_{S} (r, v)} \log_{2} (m/ \vol{E (C_{S} (\lambda ,v))}$
\end{tightlistA}
\item [2.] While \( \cost{\bdry{C_{S} (r, v)}} > \mu / (\lambda' - \lambda) \),
\begin{tightlistA}
\item [a. ] \hspace*{-0.25in} Find the vertex \( w \not \in C_{S} (r, v)
\) minimizing \( \dist{w}{C_{S}(r, v)} \) and set \( r = r +
\dist{w}{C_{S} (r, v)} \).
\end{tightlistA}
\end{tightlist2}
\end{algbox}

\begin{corollary}[Cone Cutting]\label{cor:weightedConeCutting}
Let $G = (V,E,w)$ be a connected weighted graph, let \(v\) be a vertex
in \(V\) and let $S\subseteq V$. Then for any two reals \( 0 \leq
\lambda < \lambda' \), \( \mathtt{ConeCut} (G, v,
\lambda, \lambda', S) \) returns a real \( r \in [\lambda, \lambda')
\) such that
\[
\cost{\bdry{C_S (r, v)}}  ~\leq ~
 \frac{\vol{C_S (r, v)}+\tau }{\lambda' -
\lambda}
\max \left[1,
\log_{2} \frac{m+\tau }{\vol{E (C_S (\lambda, v))}+\tau } \right] ~ ,
\]
where $m = \sizeof{E}$, and
\[
\tau ~ = ~
\left\{
\begin{array}{ll}
1, & \text{ if $\vol{E (C_S (\lambda, v))} = 0$} \\
0, & \text{ otherwise.}
\end{array}
\right.
\]
\end{corollary}

We will use two other properties of the cones $C_{S} (l,v)$:
  that we can bound their radius (Proposition~\ref{pro:radiusconcentriccone}),
  and that their removal does not increase the radius of the resulting graph
  (Proposition~\ref{pro:deleteconcentriccone}).

\begin{proposition}[Radius of Cones]\label{pro:radiusconcentriccone}
Consider a connected weighted graph \( G = (V, E, w) \), a vertex
subset \( S \subseteq V \) and let \( \psi = \max_{v \in V} \dist{v}{S}
\). Then, for every \( x \in S \),
\[
\Gradius{C_{S} (l,x)}{x}
  \leq \psi + 2 l.
\]
\end{proposition}
\begin{proof}
Let $u$ be a vertex in $C_{S} (l,x)$, and let $a_{0}, \dots , a_{k-1}$
and $b_{1}, \dots , b_{k}$ be vertices such that $a_{0} = x$,
$b_{k} = u$, $b_{i+1} \in I_{s} (a_{i})$, $(b_{i}, a_{i}) \in E$, and
\[
  \sum_{i} d(b_{i}, a_{i}) \leq l.
\]
These vertices provide a path connecting $x$ to $u$ inside $C_{S}
(l,x)$ of length at most
\[
  \sum_{i} d(b_{i}, a_{i})
+
  \sum_{i} \dist{a_{i}}{b_{i+1}}.
\]
As the first term is at most $l$, we just need to bound the second
  term by $\psi + l$.
To do this, consider the distance of each of these vertices from
  $S$.
We have the relations
\begin{align*}
\dist{b_{i+1}}{S} & = \dist{a_{i}}{S} + \dist{a_{i}}{b_{i+1}}\\
\dist{a_{i}}{S} & \geq  \dist{b_{i}}{S} - d(b_{i}, a_{i}),
\end{align*}
which imply that
\[
  \psi ~ \geq ~  \dist{b_{k}}{S}
  \geq
  \sum_{i} \dist{a_{i}}{b_{i+1}} - d(b_{i}, a_{i})
  ~ \geq ~
  \left(  \sum_{i} \dist{a_{i}}{b_{i+1}} \right) - l ~ ,
\]
as desired.
\end{proof}

In our proof, we actually use
Proposition~\ref{pro:radiusconcentriccone2} which is a slight
extension of Proposition~\ref{pro:radiusconcentriccone}.
It's proof is similar.


\begin{proposition}[Radius of Cones, II]\label{pro:radiusconcentriccone2}
Consider a connected weighted graph \( G = (V, E, w) \), a vertex
\( x_0 \in V \) and let \( \rho = \Gradius{G}{x_{0}} \). Consider a real
\( r_0 < \rho \) and let \( V_0 = B(r_0, x_0) \), \( V' = V - V_0 \)
and \( S = \BallShell{r_0}{x_0} \). Consider a vertex \( x_1 \in S \)
and let \( \psi = \rho - \Gdist{G}{x_{0}}{x_{1}} \). Then the cones
\( C_{S} (l,x_{1}) \) in the graph \( G(V') \) satisfy
\[
\Gradius{C_{S} (l,x_{1})}{x_{1}}
  \leq \psi + 2 l.
\]
\end{proposition}

\begin{proposition}[Deleting Cones]\label{pro:deleteconcentriccone}
Consider a connected weighted graph \( G = (V, E, w) \), a vertex subset \( S
\subseteq V \), a vertex \( x \in S \) and a real \( l \geq 0 \) and
let \( V' = V - C_{S}(l, x) \), \( S' = S - C_{S}(l, x) \) and \( \psi =
\max_{v \in V} \dist{v}{S} \). Then
\[
\max_{v \in V'}{\Gdist{V'}{v}{S'}} ~ \leq ~ \psi ~ .
\]
\end{proposition}
\begin{proof}
Consider some $v \in V'$.
If the shortest path from $v$ to $S$
  intersects $C_{S} (l,x)$, then
  $v \in C_{S} (l,x)$.
So, the shortest path from $v$ to $S$ in $V$
  must lie entirely in $V'$.
\end{proof}


The basic idea of $\mathtt{StarDecomp}$ is to first use
  $\mathtt{BallCut}$ to construct $V_{0}$ and then repeatedly apply
  $\mathtt{ConeCut}$ to construct $\vs{V}{1}{k}$.


\begin{algbox}\label{alg:StarDecomp}
\( \left(\{\vs{V}{0}{k},\xx,\yy \} \right) =
\mathtt{StarDecomp} (G = (V, E, w), x_{0}, \delta, \epsilon) \)
\vskip 0.1in
\begin{tightlist2}
\item [1.] Set \( \rho = \Gradius{G}{x_{0}} \); Set \( r_{0} =
\mathtt{BallCut} (G, x_{0}, \rho, \delta) \) and \( V_{0}= B
(r_{0},x_{0}) \);
\item [2.] Let \( S = \BallShell{r_0}{x_0} \);
\item [3.] Set \( G'= (V',E',w') = G (V-V_{0}) \), the weighted graph
induced by $V-V_{0}$;
\item [4.] Set \( \left(\{\vs{V}{1}{k},\xx\} \right) =
\mathtt{ConeDecomp} (G', S, \epsilon \rho / 2) \);
\item [5.] For each $i\in [1:k]$, set $y_{k}$ to be
    a vertex in $V_{0}$
    such that $(x_{k}, y_{k}) \in E$ and $y_{k}$ is on a
    shortest path from $x_{0}$ to $x_{k}$.
    Set $\yy = (\vs{y}{1}{k})$.
\end{tightlist2}
\vskip 0.1in
\( \left(\{\vs{V}{1}{k},\xx \} \right) = \mathtt{ConeDecomp}
(G, S, \Delta) \)
\vskip 0.1in
\begin{tightlist2}
\item [1.] Set $G_{0} =G $, $S_{0} = S$, and $k = 0$.

\item [2.] while $S_{k}$ is not empty
\begin{tightlistA}

\item [a. ] Set $k = k+1$; Set $x_{k}$ to be a vertex of $S_{k-1}$;
  Set \( r_{k} = \mathtt{ConeCut} (G_{k-1}, x_{k}, 0, \Delta, S_{k-1}) \)
\item [b. ]
  Set \( V_{k} = C_{S_{k-1}} (r_{k}, x_{k}) \); Set \( G_{k} = G(V -
\union_{i=1}^{k}V_{k}) \) and \( S_{k} = S_{k-1}-V_{k} \).
\end{tightlistA}
\item [3.] Set $\xx = (\vs{x}{1}{k})$.
\end{tightlist2}
\end{algbox}

\begin{proof}[Proof of Lemma~\ref{lem:stardecomposition}]
Let $\rho  = \Gradius{G}{x_{0}}$.
By setting $\delta = 1/3$, Corollary~\ref{cor:weightedBallCutting}
  guarantees \( \rho / 3 \leq
r_{0} \leq (2 / 3) \rho \).
Applying \( \Delta = \epsilon \rho / 2 \) and Propositions
  \ref{pro:radiusconcentriccone} and \ref{pro:deleteconcentriccone},
  we can bound for every $i$,
  \( r_{0} + d (x_{i},y_{i}) + r_{i} \leq \rho + 2 \Delta =
  \rho + \epsilon \rho \).
Thus \( \mathtt{StarDecomp} (G, x_{0}, 1 / 3, \epsilon) \) returns a
  \( (1 / 3, \epsilon) \)-star-decomposition with center $x_{0}$.

To bound the cost of the star-decomposition that the algorithm
produces, we use Corollaries \ref{cor:weightedBallCutting} and
\ref{cor:weightedConeCutting}.

\begin{align*}
&\cost{\bdry{V_0}}  \leq  \frac{3 \,(1+ \vol{V_0}) \log_{2} (m+1)}{\rho},
\quad \text{and}\\
&\cost{\edgbet{V_j,V - \cup_{i=0}^{j}V_i}}
    \leq  \frac{2 \, ( 1+\vol{V_j}) \log_{2} (m+1)}{\epsilon \rho}
\end{align*}
for every \( 1 \leq i \leq k \), thus
\begin{align*}
\cost{\bdry{\vs{V}{0}{k}}} & \leq ~ \sum_{j = 0}^{k}
\cost{\edgbet{V_j,V - \cup_{i=0}^{j}V_i}} \\
& \leq ~  \frac{2 \, \log_{2} (m+1)}{\epsilon \rho}
          \sum_{j = 0}^{k} ( \vol{V_j} +1) \\
& \leq ~ \frac{6 \, m \log_{2} (m+1)}{\epsilon \rho} ~ .
\end{align*}

To implement $\mathtt{StarDecomp}$ in $O (m + n\log n)$ time,
  we use a Fibonacci heap to implement steps (2) of
  $\mathtt{BallCut}$ and  $\mathtt{ConeCut}$.
If the graph is unweighted, this can be replaced by
  a breadth-first search that requires $O(m)$ time.
\end{proof}

\section{Improving the Stretch}\label{sec:Improved}

In this section, we improve the average stretch
  of the spanning tree to
  $O \left(\log^{2} n \log\log n \right)$ by
  introducing a procedure
  $\mathtt{ImpConeDecomp}$ which refines $\mathtt{ConeDecomp}$.
This new cone decomposition trades off the volume of the cone
  against the cost of edges on its boundary (similar to Seymour \cite{Seym}).
Our refined star decomposition algorithm $\mathtt{ImpStarDecomp}$ is
  identical to algorithm $\mathtt{StarDecomp}$, except that it
  calls
\[
(\{\vs{V}{1}{k},\xx \}) = \mathtt{ImpConeDecomp} (G', S,
\epsilon \rho / 2, t, \mbar)
\]
  at Step 4, where \(t\) is a positive integer that will be defined soon.


\vskip 0.1in
\begin{algbox}\label{alg:impConeDecomp}
\( \left(\{\vs{V}{1}{k},\xx \} \right) =
\mathtt{ImpConeDecomp} (G, S, \Delta, t, \mbar) \)
\vskip 0.1in
\begin{tightlist2}
\item [1.] Set \( G_{0} = G \), \( S_{0} = S \) and \( j = 0 \).

\item [2.] while \( S_{j} \) is not empty
\begin{tightlistA}

\item [a.] Set \( j = j + 1 \), set \( x_{j} \) to be a vertex of \(
S_{j - 1} \), and set \( p = t - 1 \);
\item [b.] while \( p > 0 \)
\begin{tightlistB}
\item [i.]
    \( r_j = \mathtt{ConeCut} (G_{j - 1}, x_j,
    \frac{(t - p - 1) \Delta}{t}, \frac{(t - p) \Delta}{t}, S_{j - 1}) \);
\item [ii.]
    {\bf if} \( \vol{E(C_{S_{j - 1}}(r_j, x_j))} \leq
    \frac{m}{2^{\log^{p / t} \mbar}} \) {\bf then} {\bf exit} the loop;
    {\bf else}  \( p = p - 1 \);
\end{tightlistB}
\item [c.] Set \( V_j = C_{S_{j - 1}}(r_j, x_j) \) and set \( G_{j} =
           G(V - \union_{i=1}^{j}V_{i}) \) and \( S_{j} = S_{j - 1} - V_{j} \);
\end{tightlistA}
\item [3.] Set $\xx = (\vs{x}{1}{k})$.
\end{tightlist2}
\end{algbox}
\vskip -0.1in

\begin{lemma}[Improved Low-Cost Star Decomp]\label{lem:ImpStarDecompn}
Let $G$, $x_0$ and $\epsilon$ be as in Lemma
\ref{lem:stardecomposition}, $t$ be a positive integer control
parameter, and \( \rho = \Gradius{G}{x_{0}} \). Then
\[
(\{V_0,\ldots,V_k\},\xx,\yy) =
\mathtt{ImpStarDecomp}(G, x_0, 1 / 3, \epsilon, t, \mbar) ~ ,
\]
in time $O (m + n \log n)$, returns a  $(1/3, \epsilon)$-star-decomposition of
  $G$ with center $x_0$ that satisfies
\[
\cost{\bdry{V_{0}}} ~ \leq ~ \frac{6 \, \vol{V_{0}} \log_{2} (\mbar +1)}{\rho} ~ ,
\]
and for every index $j \in \{1,2,\ldots,k\}$
   there exists $p = p(j) \in \{0,1,\ldots,t-1\}$ such that
\begin{equation} \label{eq:cost}
\cost{\edgbet{V_j,V - \cup_{i=0}^{j}V_i}}
\leq  t \cdot \frac{4 \, \vol{V_j} \log^{(p + 1) / t} (\mbar+1)}
                    {\epsilon \rho} ~ ,
\end{equation}
and unless \( p = 0 \),
\begin{equation} \label{eq:volume}
\vol{E(V_j)} ~ \leq ~ \frac{m}{2^{\log^{p/t} \mbar}} ~ .
\end{equation}

\end{lemma}
\begin{proof}
In what follows, we call \( p(j) \) the \emph{index-mapping} of
the vertex set \( V_j \).
We begin our proof by observing that \( 0 \leq r_j < \epsilon \rho
/ 2 \) for every \( 1 \leq j \leq k \).
 We can then show that \( \{\vs{V}{0}{k}\} \)
  is a \( (1 / 3, \epsilon) \)-star decomposition as we did in the proof of
Lemma \ref{lem:stardecomposition}.

We now bound the cost of the decomposition.
Clearly, the bound on \( \cost{\bdry{V_{0}}} \) remains
  unchanged from that proved in Lemma
  \ref{lem:stardecomposition}, but here we bound
  $\vol{V_{0}} + 1$ by $2 \vol{V_{0}}$.

Below we will use \( \Delta = \epsilon \rho / 2 \) as specified in the algorithm.

Fix an index \( j \in \{1,2,\ldots,k\} \), and let \( p = p(j) \) be
the final value of variable \(p\) in the loop above (that is, the
value of \(p\) when the execution left the loop while constructing \(
V_j \)). Observe that \( p \in \{0,1,\ldots,t-1\} \), and that unless the
loop is aborted due to \( p = 0 \), we have \( \vol{E(V_j)} \leq
\frac{m}{2^{\log^{p/t} \mbar}} \) and inequality \eqref{eq:volume} holds.

For inequality \eqref{eq:cost}, we split the discussion to two
cases. First, consider the case \( p = t - 1 \).
Then the inequality
  \( \cost{\edgbet{V_j, V - \cup_{i=0}^{j}V_i}} \leq
  (\vol{V_j}+1)  \log (\mbar +1) (t/\Delta) \) follows directly from Corollary
  \ref{cor:weightedConeCutting}, and inequality \eqref{eq:cost} holds.

Second, consider the case \( p < t - 1 \) and let \( r_{j}' \) be the
value of the variable \( r_j \) at the beginning of the last
iteration of the loop (before the last invocation of Algorithm
\texttt{ConeCut}). In this case, observe that at the beginning of the
last iteration, \( \vol{E(C_{S_{j - 1}}(r_{j}', x_j))} >
\frac{m}{2^{\log^{(p + 1) / t} \mbar}} \) (as otherwise the loop
would have been aborted in the previous iteration).
By Corollary \ref{cor:weightedConeCutting},
\[
 \cost{\edgbet{V_j, V - \cup_{i=0}^{j}V_i}}
 ~\leq ~
  \frac{\vol{V_j}}{\Delta / t} ~ \times
\max \left[1,\log_{2}{\left( \frac{m}{\vol{E \left(
C_{S_{j - 1}} \left( \frac{(t - p - 1) \Delta}{t}, x_j \right) \right)}}
\right)} \right],
\]
where \( V_j = C_{S_{j - 1}}(r_j, x_j) \).
Since
\[
\frac{(t - p - 2)
\Delta}{t} \leq r_{j}' < \frac{(t - p - 1) \Delta}{t},
\]
it follows that
\[
\vol{E \left( C_{S_{j - 1}} \left( \frac{(t - p - 1) \Delta}{t}, x_j
\right) \right)}
~ \geq ~ \vol{E(C_{S_{j - 1}}(r_{j}', x_j))} ~ > ~
\frac{m}{2^{\log^{(p + 1) / t} \mbar}} ~ .
\]
Therefore
\[
\log^{(p + 1) / t} \mbar ~ \geq ~
\max \left[1,\log_{2}{\left( \frac{m}{\vol{E \left(
C_{S_{j - 1}} \left( \frac{(t - p - 1) \Delta}{t}, x_j \right) \right)}}
\right)} \right] ~ ,
\]
and
\[
\cost{\edgbet{V_j, V - \cup_{i=0}^{j}V_i}}  ~ \leq ~ \frac{\vol{V_j}
\log^{(p + 1) / t} \mbar}{\Delta / t}
 ~ = ~ t \cdot \frac{2 \, \vol{V_j} \log^{(p + 1) / t}
\mbar}{\epsilon \rho} ~ .
\]
\end{proof}

Our improved algorithm \( \mathtt{ImpLowStretchTree} (G, x_{0}, t, \mbar) \),
is identical to \texttt{LowStretchTree} except that in Step 3 it calls
$\mathtt{ImpStarDecomp} (\Contracted{G}, x_{0}, 1/3, \beta, t,
\mbar),$ and in Step 5 it calls \AvoidOverfull{}
$\mathtt{ImpLowStretchTree} (G (V_{i}), x_{i}, t, \mbar)$. We set $t =
\log\log n$ throughout the execution of the algorithm.

\begin{theorem}[Lower--Stretch Spanning Tree] \label{thm:main2}
Let \( G = (V, E, w) \) be a connected weighted graph and let \(x_0\)
be a vertex in \(V\). Then
\[
T = \mathtt{ImpLowStretchTree} (G, x_{0}, t, \mbar) ~ ,
\]
in time \( O (\widehat{m} \log \widehat{n} + \widehat{n} \log^2
\widehat{n}) \), returns a spanning tree of \(G\) satisfying
\[
\Gradius{T}{x_0} \leq 2 \sqrt{e} \cdot  \Gradius{G}{x}
\]
and
\[
\avestretch{T}{E} = O \left(\log^{2} \widehat{n} \log\log \widehat{n}
\right) ~ .
\]
\end{theorem}
\begin{proof}

The bound on the radius of the tree remains unchanged from that proved
in Theorem~\ref{thm:main}.

We begin by defining a system of notations for the recursive process,
assigning for every graph \( G = (V, E) \) input to some recursive
invocation of Algorithm \texttt{ImpLowStretchTree}, a sequence \(
\GraphSequence{G} \) of non-negative integers. This is done as
follows. If \(G\) is the original graph input to the first invocation
of the recursive algorithm, than \( \GraphSequence{G} \) is
empty. Assuming that the halt condition of the recursion is not
satisfied for \(G\), the algorithm continues and some of the edges in
\(E\) are contracted. Let \( \Contracted{G} = (\Contracted{V},
\Contracted{E}) \) be the resulting graph. (Recall that we refer to \(
\Contracted{G} \) as the edge-contracted graph.) Let \( \{
\Contracted{V}_0, \Contracted{V}_1, \dots , \Contracted{V}_k \} \) be
the star decomposition of \( \Contracted{G} \). Let \( V_j \in V \) be
the preimage under edge contraction of \( \Contracted{V}_j \in
\Contracted{V} \) for every \( 0 \leq j \leq k \). The graph \( G(V_j)
\) is assigned the sequence \( \GraphSequence{G({V}_{j})} =
\GraphSequence{G} \cdot j \). Note that \( | \GraphSequence{G} | = h
\) implies that the graph \(G\) is input to the recursive algorithm on
recursion level \(h\). We warn the reader that the edge-contracted
graph obtained from a graph assigned with the sequence \(\sigma\) may
have fewer edges than the edge-contracted graph obtained from the
graph assigned with the sequence \( \sigma \cdot j \), because the
latter may contain edges that were contracted out in the former.

We say that the edge \(e\) is \emph{present} at recursion level \(h\)
if \(e\) is an edge in \( \Contracted{G} \) which is the
edge-contracted graph obtained from some graph \(G\) with \( |
\GraphSequence{G} | = h \) (that is, it was not contracted out). An
edge $e$ \emph{appears} at the first level \(h\) at which it is
present, and it \emph{disappears} at the level at which it is present
and its endpoints are separated by the star decomposition. If an edge
appears at recursion level \(h\), then a path connecting its endpoints
was contracted on every recursion level smaller than \(h\), and no
such path will be contracted on any recursion level greater than
\(h\). Moreover, an edge is never present at a level after it
disappears. We define \( h(e) \) and \( h'(e) \) to be the recursion
levels at which the edge \(e\) appears and disappears, respectively.

For every edge $e$ and every recursion level $i$ at which it is
present, we let $U (e,i)$ denote the set of vertices $\Contracted{V}$
of the edge-contracted graph containing its endpoints.
If $h (e) \leq i < h' (e)$, then we let $W (e,i)$ denote the set of vertices
  $\Contracted{V}_{j}$ output by \texttt{ImpStarDecomposition}
  that contains the endpoints of $e$.

Recall that \( p(j) \) denote the index-mapping of the vertex set \(
V_j \) in the star decomposition. For each index \( i \in
\{0,1,\ldots,t-1\} \), let \( I_i = \{j \in \{1,2,\ldots,k\} \mid p(j)
= i\} \). For a vertex subset \( U \subseteq V \), let \( \AS{U} \) denote the
average stretch that the algorithm guarantees for the edges of
$E(U)$. Let \( \TS{U} = \AS{U} \cdot | E(U) | \). Then by Lemma
\ref{lem:ImpStarDecompn}, the following recursive formula applies.
\begin{eqnarray}
\TS{V} & \leq & \left( \sum_{j = 0}^k \TS{\Contracted{V}_j} \right)
\nonumber \\
& + & 4 \sqrt{e} \times \left( 6 \, \log (\widehat{m} + 1) \cdot
\vol{\Contracted{V}_0} ~ + ~ 4 \, \frac{t}{\beta} \sum_{p = 0}^{t - 1}
\log^{(p + 1) / t} ( \mbar+1) \sum_{j \in I_p} \vol{\Contracted{V}_j}
\right) \nonumber \\
& + & \left( \sum_{e \in E - \Contracted{E}}{\stretch{T}{e}} \right)
\nonumber \\
& = & \left( \sum_{j = 0}^k \TS{V_j} \right) \nonumber \\
& + & 4 \sqrt{e} \times \left( 6 \, \log (\widehat{m} + 1) \cdot
\vol{\Contracted{V}_0} ~ + ~ 4 \, \frac{t}{\beta} \sum_{p = 0}^{t - 1}
\log^{(p + 1) / t} ( \mbar+1) \sum_{j \in I_p} \vol{\Contracted{V}_j}
\right) ~ , \label{eq:TS}
\end{eqnarray}
where we recall $\beta = \epsilon = (2 \log_{4/3} (n + 32))^{-1}$.

For every edge \(e\) and for every \( h(e) \leq i < h'(e) \), let
\( \pi_i(e) \) denote the index-mapping of the component \(
W(e, i) \) in the invocation of Algorithm \texttt{ImpConeDecomp}
on recursion level \(i\). For every index \( p \in \{ 0, \dots , t - 1
\} \), define the variable \( l_p(e) \) as follows
\[
l_p(e) ~ = ~ \left| \left\{ h(e) \leq i < h'(e) \mid \pi_i(e) = p
\right\} \right| ~ .
\]

For a fixed edge \(e\) and an index
  \( p \in \{ 0, \dots , t - 1 \} \),
  every \( h(e) \leq i < h'(e) \) such that \( \pi_i(e) = p \)
   reflects a contribution of
  \( O (t / \beta) \cdot \log^{(p + 1) / t} (\mbar + 1) \)
  to the right term in
  \eqref{eq:TS}.
Summing $p$ over $\setof{0, 1, \ldots , t-1}$, we obtain
\[
\sum _{p=0}^{t-1} O (t / \beta) l_{p} (e)\log^{(p + 1) / t} (\mbar + 1).
\]
In a few moments, we will prove that
\begin{equation}\label{eqn:sumLp}
\sum _{e} \sum _{p=0}^{t-1} l_{p} (e)\log^{p  / t} (\mbar + 1)
  \leq
O (\mbar \log_{2} \mbar),
\end{equation}
which implies that the sum of the contributions of all edges $e$ in levels
  \( h(e) \leq i < h'(e) \) to the right term in
  \eqref{eq:TS}
 is
\[
 O \left(
\frac{t}{\beta} \cdot \mbar \log^{1 + 1 / t} \mbar \right)
=
 O \left(
t \cdot \mbar \log^{2 + 1 / t} \mbar \right)
\]

As  \( \vol{V_j} \) counts the internal edges of \(V_j \) as well as
  its boundary edges, we must also account for the contribution of
  each edge $e$ at level $h' (e)$.
At this level, it will be counted twice---once in each component containing
  one of its endpoints.
Thus, at this stage, it contributes a factor of at most
  \( O ((t/\beta ) \cdot \log \mbar) \) to the sum \( \TS{V} \).
Therefore all edges
  \( e \in E \)
contribute an additional factor of
 \( O (t \cdot \mbar \log^2 \mbar) \).
Summing over all the edges, we find that all the contributions to
  the right term
  in
  \eqref{eq:TS} sum to at most
\[
 O \left(
t \cdot \mbar \log^{2 + 1 / t} \mbar \right) .
\]

Also, every \( h(e) \leq i < h'(e) \) such that the edge \(e\) belongs
to the central component \( \Contracted{V}_0 \) of the star
decomposition, reflects a contribution of \( O (\log \mbar) \) to the
left term in \eqref{eq:TS}. Since there are at most \( O (\log \mbar) \) such
\(i\)s, it follows that the contribution of the left term in
\eqref{eq:TS} to \( \TS{V} \) sums up to an additive term of \( O
(\log^2 \mbar) \) for every single edge, and \( (\mbar \log^2 \mbar) \)
for all edges.

It follows that \( \TS{V} = O (t \cdot \mbar \log^{2 + 1 / t}
\mbar) \). This is optimized by setting \( t = \log\log \mbar \),
obtaining the desired upper bound of \( O (\log^{2} \widehat{n} \cdot \log\log
\widehat{n}) \) on the average stretch \( \AS{V} \) guaranteed by Algorithm
\texttt{ImpLowStretchTree}.

We now return to the proof of \eqref{eqn:sumLp}.
We first note that  \( l_0(e) \) is at
  most \( O (\log \mbar) \) for every edge \(e\).
We then observe that for each index
  \( p > 0 \) and each \( h(e) \leq i < h'(e) \) such that \( \pi_i(e) = p  \),
  $\vol{E(U (e,i))}/\vol{E(W (e,i))}$
  is at least
  \( 2^{\log^{p / t} \mbar} \) (by Lemma
  \ref{lem:ImpStarDecompn}, \eqref{eq:volume}).
For $h (e) \leq  i < h' (e)$, let
  $g_{i} (e) = \vol{E(U (e,i+1))} / \vol{E(W (e,i))}$.
We then have
\[
\prod _{1 \leq p \leq t-1}
\left( 2^{\log^{p / t} \mbar} \right)^{l_p(e)} ~
\leq ~ \mbar \prod _{h (e) \leq  i < h' (e)} g_{i} (e) ~ ,
\]
hence \( \sum _{p=1}^{t-1} l_p(e) \log^{p / t} \mbar \leq \log
\mbar + \sum _{h (e) \leq  i < h' (e)} \log g_{i} (e)
 \).
We will next prove that
\begin{equation}\label{eqn:sumgis}
\sum _{e} \sum _{h (e) \leq  i < h' (e)} \log g_{i} (e)
\leq \mbar \log \mbar,
\end{equation}
which implies \eqref{eqn:sumLp}.

Let $E_{i}$ denote the set of edges present at recursion level $i$.
For every edge $e \in E_{i}$ such that $i < h' (e)$, we have
\[
  \sum _{e' \in E(W (e,i))} g_{i} (e') = g_{i} (e) \vol{E(W (e,i))}
 = \vol{E(U (e,i+1))},
\]
and so
$  \sum _{e \in E_{i} : i < h' (e) } g_{i} (e) = \vol{E_{i+1}}.$
As each edge is present in at most $O (\log \mbar)$ recursion depths,
$\sum _{i} \vol{E_{i}} \leq \mbar \log \mbar$, which proves \eqref{eqn:sumgis}.
\end{proof}

\section{Conclusion}\label{sec:conclusion}

At the beginning of the paper, we pointed out that the definition
  of stretch used in this paper differs slightly from that used
  by Alon, Karp, Peleg and West~\cite{AKPW}.
If one is willing to accept a longer running time, then this problem
  is easily remedied as shown in Subsection \ref{subsec:applications}.
If one is willing to accept a bound of $O (\log ^{3} n)$ on the stretch,
  then one can extend our analysis to show that the natural randomized
  variant \texttt{LowStretchTree},  in which
  one chooses the radii of the balls and cones at random, works.

A natural open question is whether one can improve the stretch bound
   from $O\left(\log^{2} n\log\log n \right)$ to $O
  (\log n)$.
Algorithmically, it is also desirable to improve the running time of
  the algorithm to $O (m\log n)$.
If we can successfully achieve both improvements, then
   we can use the Spielman-Teng solver to solve planar diagonally dominant
   linear systems in $O (n\log n \log (1/\epsilon ))$ time.

As the average stretch\footnote{In the context of the optimization
problem of finding a spanning tree with the lowest average stretch,
the stretch is defined as in \cite{AKPW}.} of any spanning tree in a weighted connected graph is $\Omega (1)$, our low-stretch tree algorithm also provides
  an $O\left(\log^{2} n\log\log n \right)$-approximation
  to the optimization problem of finding
  the spanning tree with the lowest average stretch.
It remains open (a) whether our algorithm has a better approximation
  ratio and (b) whether one can in polynomial time find a spanning tree
  with better approximation ratio, e.g., $O (\log n)$ or even $O (1)$.

\section{Acknowledgments}\label{sec:ack}
We are indebted to David Peleg, who was offered co-authorship on this paper.

\end{document}

%% file: prestuff.tex
\newtheorem{theorem}{Theorem}[section]
\newtheorem{corollary}[theorem]{Corollary}
\newtheorem{lemma}[theorem]{Lemma}

\newtheorem{proposition}[theorem]{Proposition}

\theoremstyle{definition}
\newtheorem{definition}[theorem]{Definition}
\theoremstyle{plain}

